\input{aipcheck}

\documentclass[
    ,final            
  ]
  {aipproc}

\layoutstyle{6x9}

\begin{document}

\title{AnDY : Overview and Plans \\ Feasibility Test of Large Rapidity Drell-Yan Production at RHIC}


\author{Chris Perkins}{
  address={UC Berkeley/Space Sciences Lab, Stony Brook University}
}

\begin{abstract}
Measuring the single transverse spin asymmetry $A_N$ for large $x_F$ Drell-Yan production provides the most robust test of our current theoretical understanding of transverse proton spin structure.
This includes a measurement of the predicted sign change of the Sivers function in comparison to that found in SIDIS data.
The first year of a feasibility test has been completed to define the requirements for measuring large $x_F$ Drell-Yan production at RHIC.
The impact of colliding beams at a third IR at RHIC was found to be negligible and a rich jet-triggered data sample was acquired by AnDY in RHIC Run 11.
Plans for subsequent runs are currently underway.

\end{abstract}

\maketitle


\section{Large Rapidity Drell-Yan at a Collider}

Concurrent with the measurement of large transverse single spin asymmetries reported at RHIC \cite{RHIC} and HERMES \cite{HERMES}, theories were further developed to attempt to explain these phenomena in terms of Transverse Momentum Dependent (TMD) distribution and fragmentation functions.
In these theories, a gauge link that appears as an attractive final state interaction in Semi-Inclusive Deep Inelastic Scattering becomes a repulsive initial state interaction in the Drell-Yan process, leading to the prediction that the sign of the Sivers function will be opposite in DIS as compared to Drell-Yan \cite{SignChange}.
The measurement of this sign change is a robust test of our current understanding of transverse spin physics and TMD factorization.
The importance of this measurement was recognized in DOE Milestone HP13 with the charge to test this prediction by 2015.
While theoretical factorization has encountered problems in other processes, factorization is still thought to be robust in Drell-Yan because of the simple color structure.
Measurement of the analyzing power ($A_N$) for jets in p+p collisions is a prerequisite to this Drell-Yan measurement and can resolve a sign ambiguity in theory \cite{ANjet}.

Separating the Drell-Yan signal from the background at large rapidities at a collider is no easy task and the experimental requirements for such a measurement cannot be fully determined from simulations alone.
A feasibility test has begun at RHIC to define the requirements which can then be used in future major forward upgrades at STAR and PHENIX.
The most important question is whether the measurement can be made with calorimetry alone or if tracking is required.
Establishing the requirements for a large $x_F$ Drell Yan production experiment will also lead to future avenues that provide the most robust interconnections between low-x probed at RHIC and low-x probed at eRHIC.

The proposed feasibility test would be timely in that it could run concurrently with the STAR and PHENIX W programs (2012-2013) and could complete the DOE Milestone HP13 by 2015.
Severe space constraints at STAR and PHENIX would require major changes in the forward direction to do a Drell Yan experiment making the feasibility test at a third interaction point at RHIC a natural solution.
A Letter of Intent (LOI) was submitted to the Program Advisory Committee (PAC) at BNL in May 2010 \cite{PAC2010} with the intent to start such a feasibility test at IP2 at RHIC in Run 11 and was supported by the PAC for this run.
After RHIC Run 11, a further proposal was submitted and also supported by the PAC to continue the planned program in Run 12 \cite{PAC2011}.

The primary goals for the Drell-Yan measurement at IP2 at RHIC are:  1) Establish that large-$x_F$ low-mass dileptons from the Drell Yan process can be discriminated from the background in $\sqrt{s}$ = 500 GeV polarized p+p collisions.  2) Provide sufficient statistical precision for the analyzing power $A_N$ for Drell Yan production to test the theoretical prediction of a sign change compared to transverse single spin asymmetries for SIDIS.

Looking at the $e^+$$e^-$ Drell-Yan process provides the possibility to perform the measurement with calorimetry alone.
A larger signal can be obtained at RHIC by using $\sqrt{s} = 500$ GeV p+p collisions (compared to $\sqrt{s} = 200$ GeV) but a high luminosity will be required ($\sim$ 100 $pb^{-1}$) along with a very forward detector with good background separation.
The detector must be able to accurately separate electrons from hadrons and charged particles from neutrals.
As seen in the PAC proposal, PYTHIA + GEANT simulations were performed and indicate that reasonable efficiency can be obtained for large $x_F$ Drell-Yan with the equipment described.

\section{RHIC Run 11 Accomplishments}
The AnDY Experiment had three goals for RHIC Run 11 : 
1) Establish the impact on STAR and PHENIX of colliding beams at three interaction regions at RHIC. 
2) Calibrate the hadron calorimeter's absolute energy scale.
3) Measure hadronic background to benchmark Monte Carlo studies further.
With an integrated luminosity of $\sim$ 10 $pb^{-1}$ and beam polarization of 50$\%$, the analyzing power for jets could also be determined.

\begin{figure}
  \includegraphics[width=0.29\textwidth]{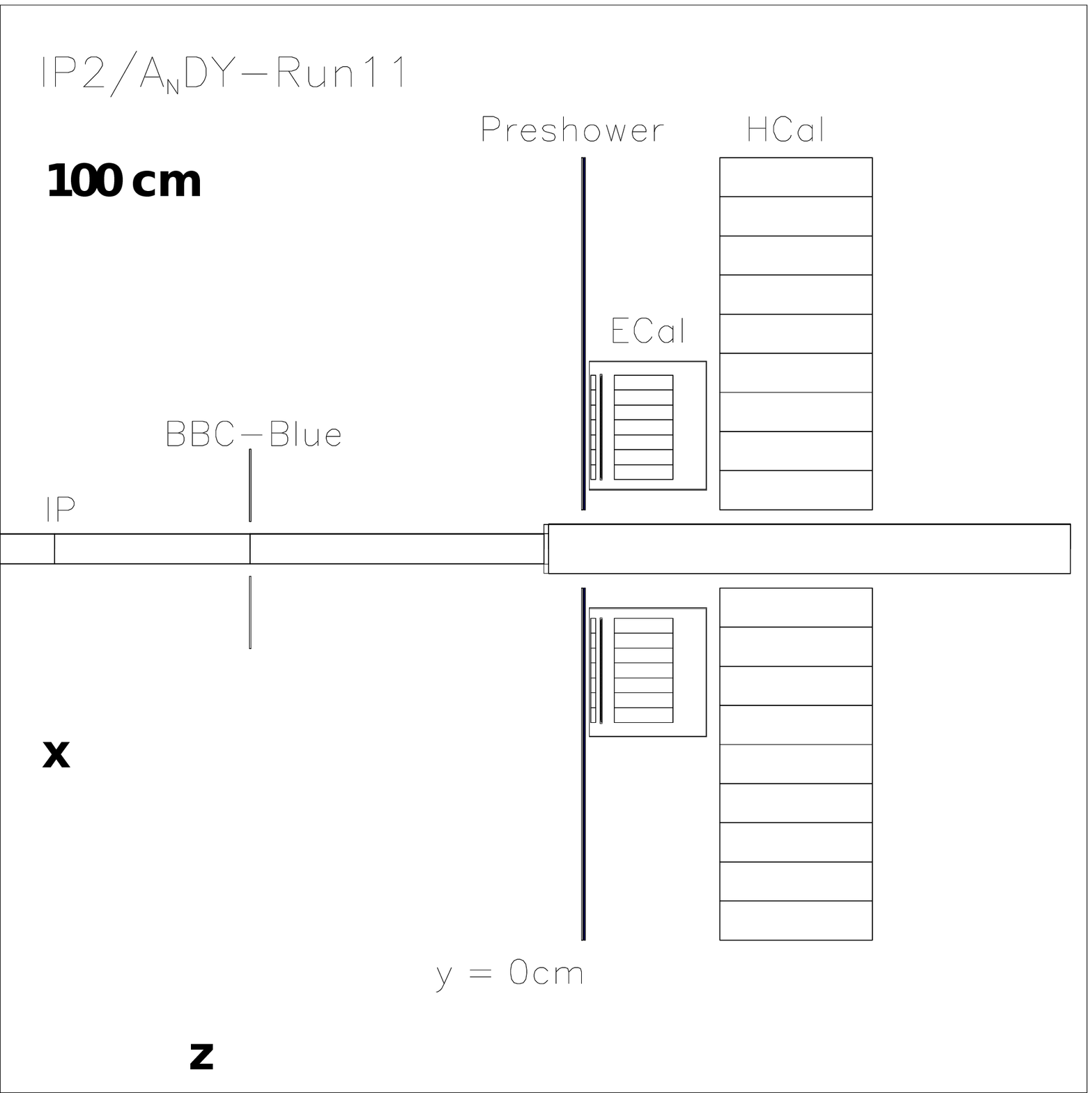}
  \includegraphics[width=0.31\textwidth]{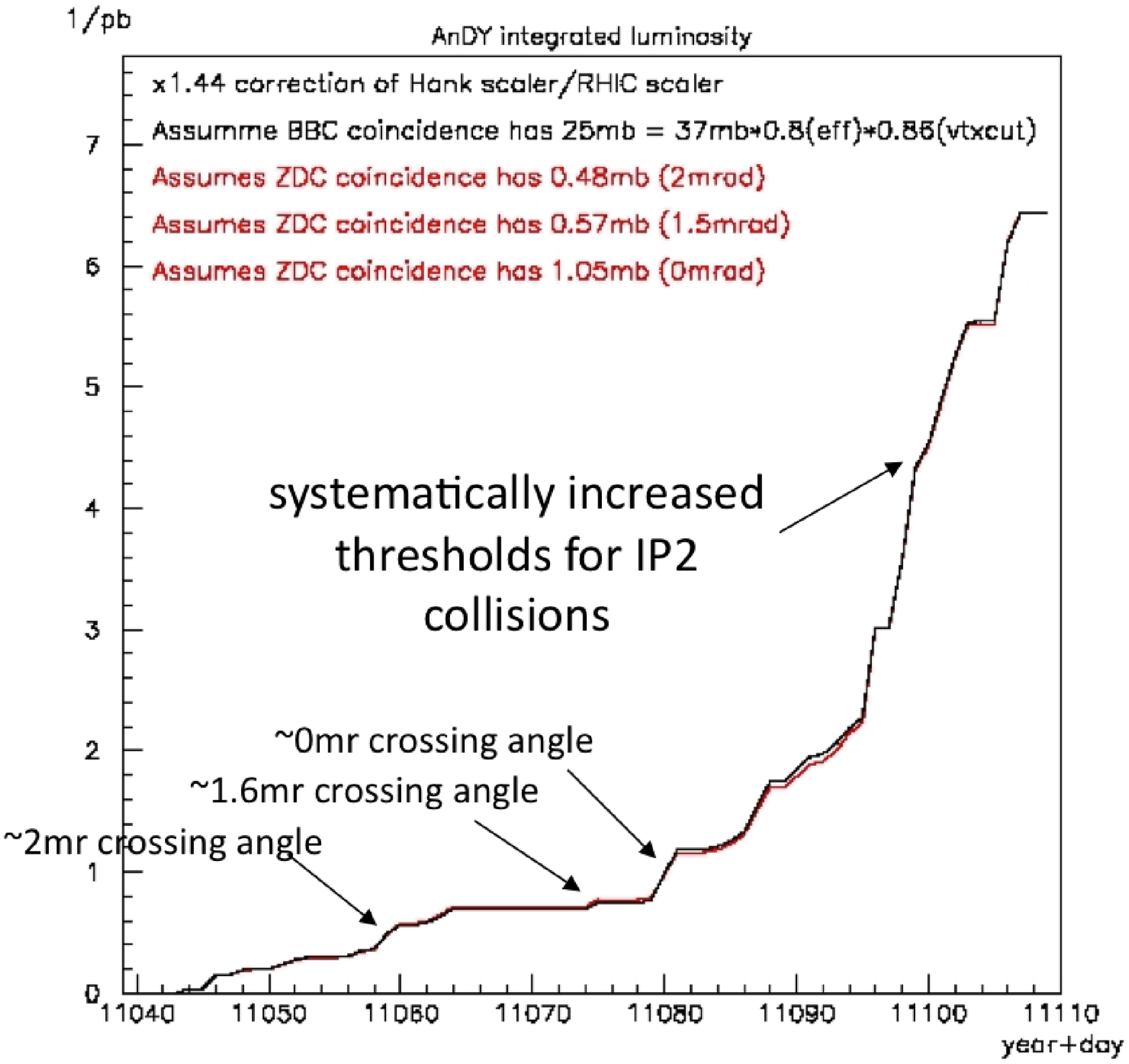}
  \includegraphics[width=0.33\textwidth]{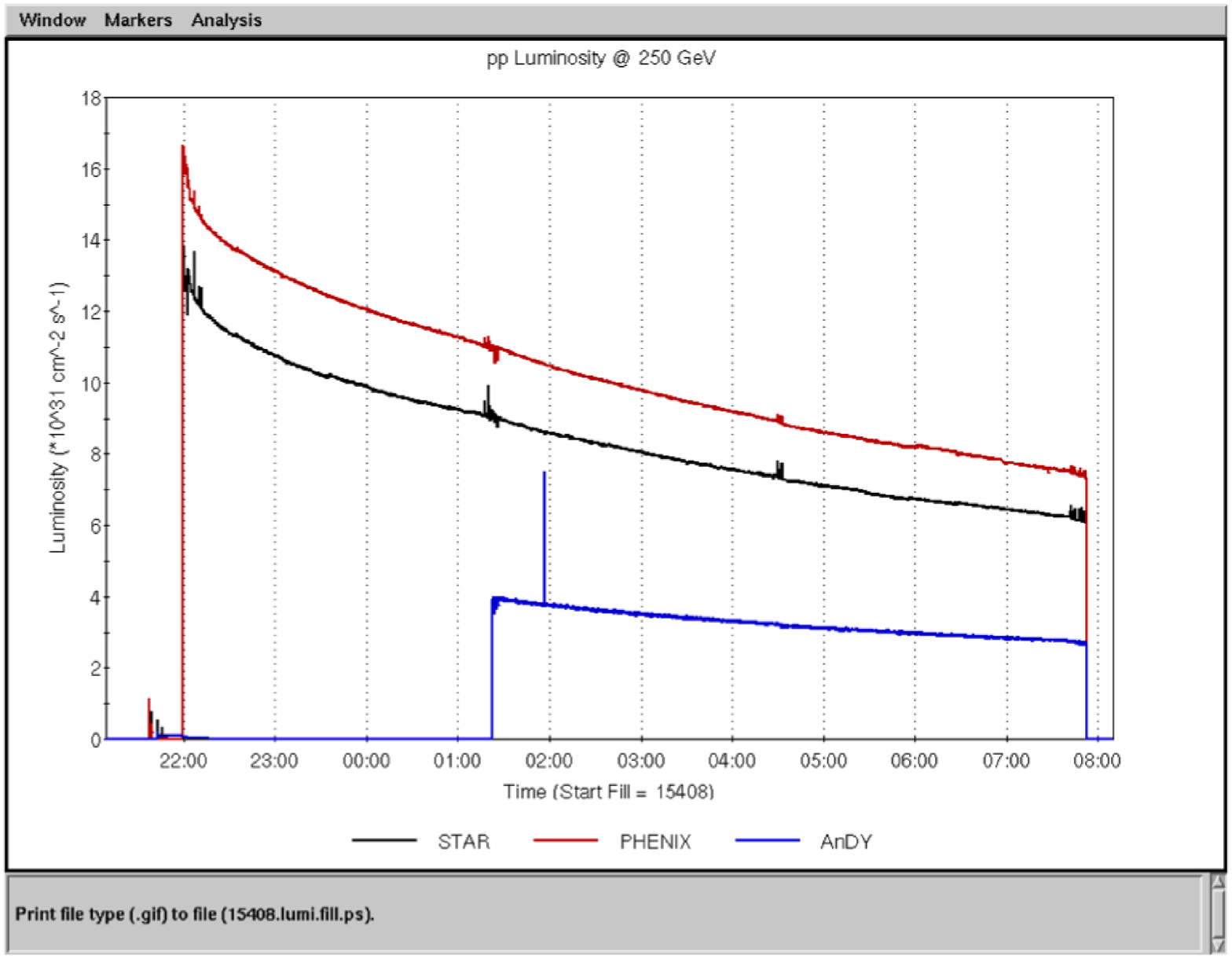}
  \caption{
(Left) Top-view of AnDY configuration for RHIC Run 11.
(Middle) AnDY Integrated Luminosity versus day in RHIC Run 11 p+p at $\sqrt{s}=500$ GeV.
(Right) Luminosity for one example fill (Fill 15408) during RHIC Run 11.
   }
  \label{andy_run11}
\end{figure}

In the initial LOI, a suite of detectors was proposed for RHIC Run 11 to accomplish these goals, all of which were successfully assembled and installed prior to this run.
A top view of the AnDY Experiment in RHIC Run 11 is shown in Figure~\ref{andy_run11}.

The Hadron Calorimeter (HCal) used existing cells each composed of an array of scintillating fibers embedded in lead.
The HCal cells were previously used in E864 at the AGS at BNL.
Beam-Beam Counters (BBC) previously used at PHOBOS were installed on either side of the interaction point to allow for a basic collision trigger.
Zero-Degree Calorimeters (ZDC) and ZDC Shower Maximum Detectors (ZDC-SMD), also used previously at RHIC, were also installed on either side of the interaction point and allowed for local polarimetry measurements and other monitoring (Not shown in Figure~\ref{andy_run11}).
The only detector component used at AnDY in Run 11 that had not been borrowed or obtained from a previous experiment was the Preshower Detector.
In addition, two small, modular Electromagnetic Calorimeters (ECal) were placed symmetrically on either side of the beam.
Approximately 120 lead-glass cells were borrowed from the BigCal detector at Jefferson Lab to build these detectors, forming two 7x7 arrays.

Readout and trigger electronics based on the STAR system were also obtained, built, and installed.
A full trigger/DAQ system was built using custom designed electronics that can trigger on complex patterns of any combination of the detectors described above.
The trigger system can look at every RHIC crossing ($\sim$9.4 MHz) for a trigger, leading to nearly zero deadtime.
A suite of triggers was developed that was capable of triggering on LEDs, Cosmic-rays, minimum bias collisions, $\pi^0$ in ECal, and jets in HCal.

The integrated luminosity at AnDY for p+p $\sqrt{s}$ = 500 GeV collisions over the course of Run 11 is shown in Figure~\ref{andy_run11}.
Several improvements were made over the course of the run by the Collider-Accelerator Department that helped increase the integrated luminosity dramatically.
The beam crossing angle was reduced from $\sim$2 mrad to 0 mrad.
Early in the run, the beams were only steered into collisions at AnDY after the bunch intensity had fallen below a fairly low limit, leading to relatively short collision periods for AnDY.
As the run progressed, the bunch intensity threshold was gradually raised to assess the impact on luminosities at STAR and PHENIX, leading to increased luminosity at AnDY.
A reduction in $\beta^{*}$ at IP2 can further increase luminosity by a factor of $\sim$ 2.

\begin{figure}
  \includegraphics[width=0.23\textwidth]{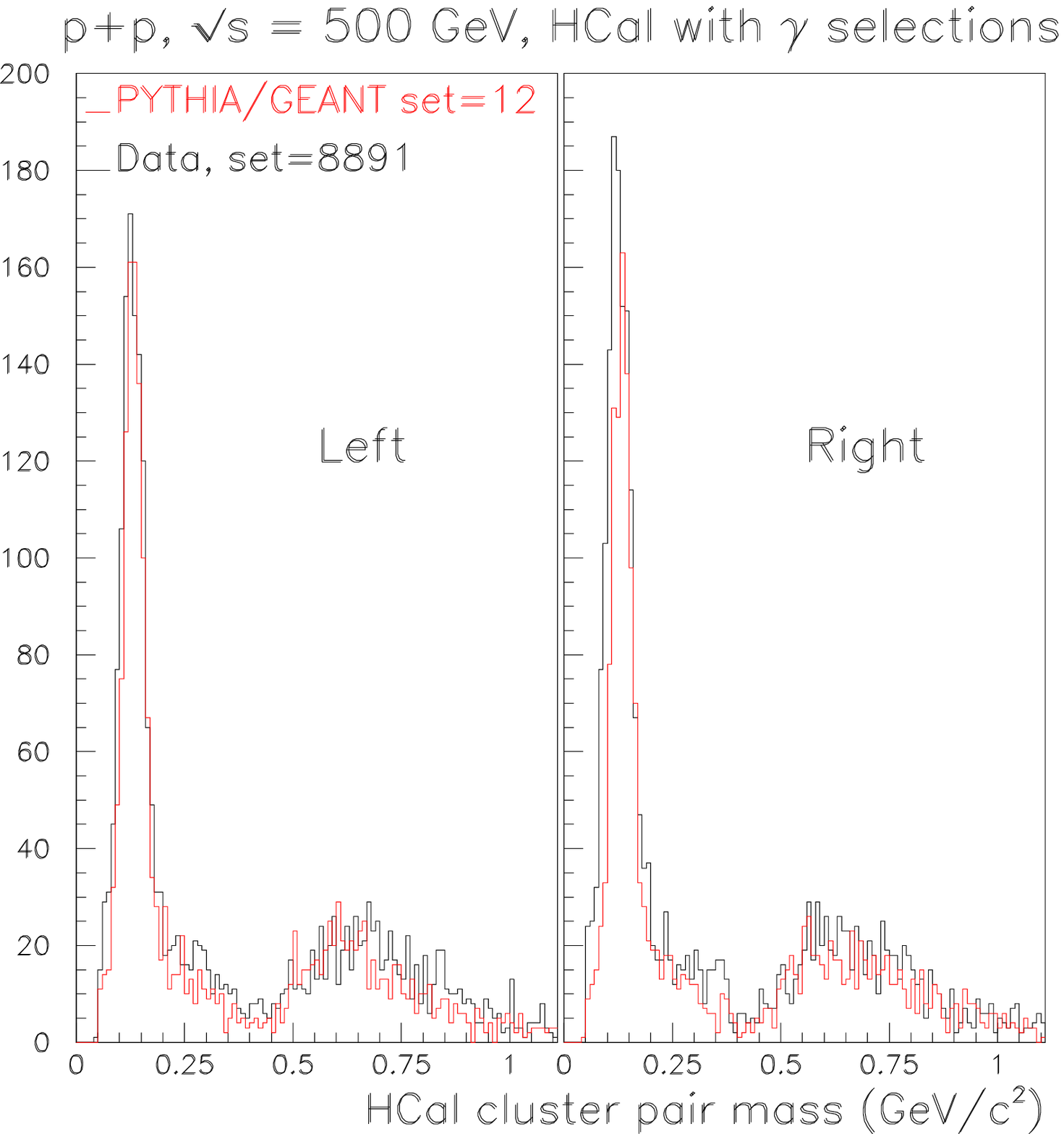}
  \includegraphics[width=0.26\textwidth]{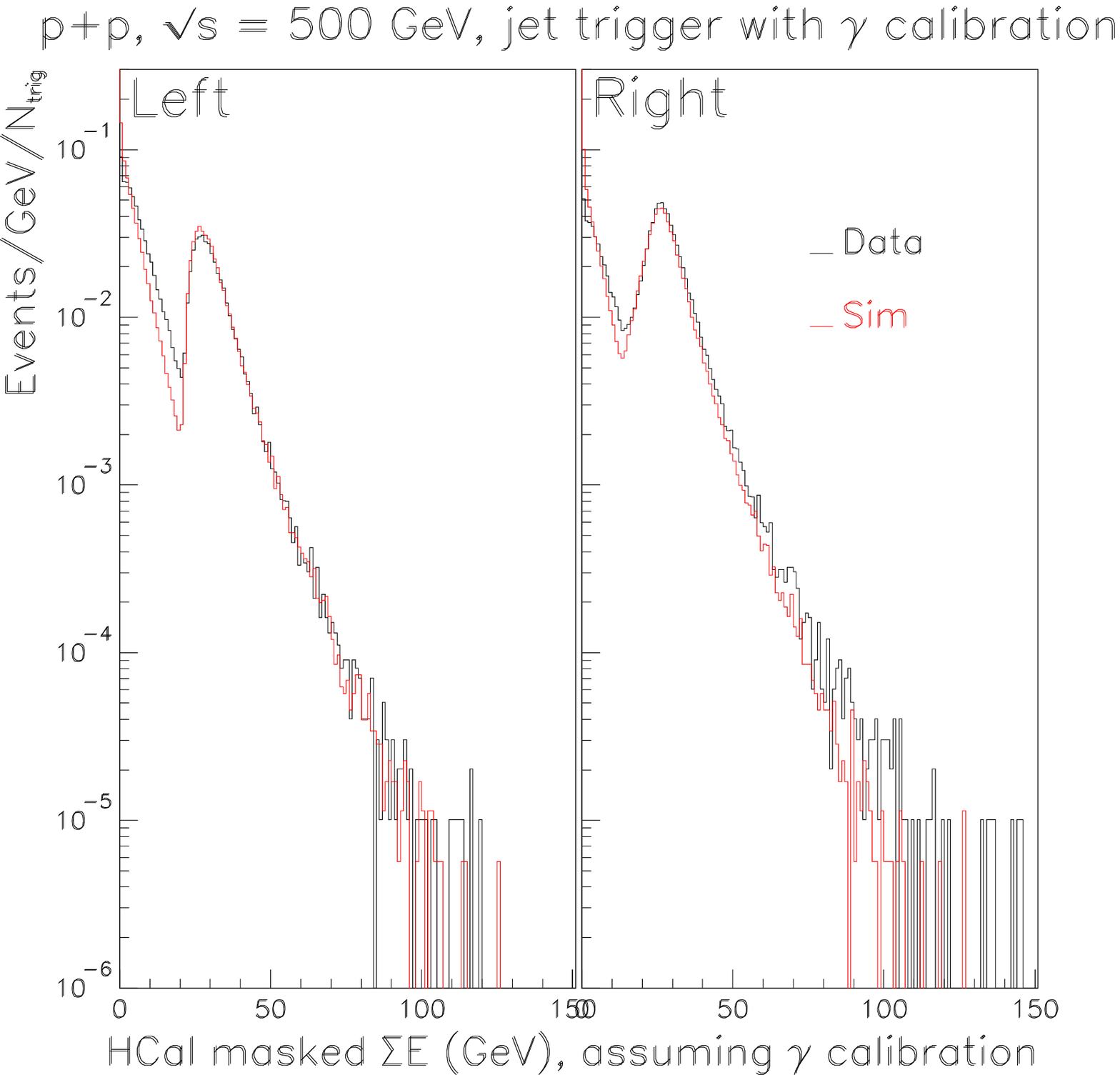}
  \caption{
Data and Simulation for
(Left) Reconstructed invariant mass in HCal modules for clusters subjected to photon-like requirements from a minimum bias trigger.
(Right) Summed energy from HCal modules from the Run 11 jet trigger after initial calibrations.
   }
  \label{andy_run11_hcal}
\end{figure}

An example of the luminosities at PHENIX, STAR, and AnDY over the course of one fill is shown in Figure~\ref{andy_run11}.
The beams were steered into collisions at AnDY for this particular fill when the threshold of < 1.4x$10^{11}$ ions/bunch was reached.
The small deviations in the STAR and PHENIX luminosities around 01:30 and 04:30 were due to polarization measurements.
Some fills were not as clean as the example shown but after a procedure was established for bringing the beams into collisions at AnDY, most fills were approximately as smooth as the one shown.
When the procedure was followed, it can be seen that the impact of collisions at AnDY on the integrated luminosities at STAR and PHENIX is small.
In meetings subsequent to DIS 2011, the Collider-Accelerator Department has stated that integrated luminosities of $\sim 10 pb^{-1}$ per week can be delivered to AnDY with minimal impact on STAR and PHENIX.

During RHIC Run 11, more than 750 million jet triggered events were collected.
The HCal was calibrated by reconstruction of $\pi^0$ from clusters subjected to photon-like requirements.
The reconstructed invariant mass is shown in Figure~\ref{andy_run11_hcal}.
An association analysis of simulations confirms the peak is from neutral pions.
This peak establishes the energy scale of HCal to $\sim5$\%.
Corrections are still required to reconstruct the incident energies of hadrons.
Reconstruction of full jets using these calibrations is underway, as shown by the summed energy distribution in Figure~\ref{andy_run11_hcal}.

\section{RHIC Runs 12 and 13 Plans}

As described in the LOI, a loan agreement for the BigCal Detector at Jefferson Lab has been made and the transfer of lead-glass cells and photomultiplier tubes to BNL was completed in July 2011.
A sample of $\sim$100 $pb^{-1}$ will be obtained with this expanded electromagnetic calorimeter with the goals of measuring J/$\Psi$, $\Upsilon$, and the continuum between them and attempting to measure $A_N$(DY) without a magnet.
After this, the split-dipole magnet from the PHOBOS experiment will be installed along with fiber trackers.
Another $\sim$100 $pb^{-1}$ sample will be obtained to see the improvement in signal-to-background for $A_N$(DY) when tracking through a split-dipole magnet is used.

\section{Conclusions and Outlook}

Establishing the requirements for a large $x_F$ Drell-Yan production experiment will provide the most robust test of theory for transverse spin and lead to future avenues that provide the most robust interconnections between low-x probed at RHIC and low-x proved at eRHIC.
A feasibility test is underway to provide these requirements and collected a significant sample of jet-triggered data in RHIC Run 11.
The impact on other experiments at RHIC was found to be negligible when colliding beams at a third IR.
Plans for the completion of the detector required for the feasibility test are underway with significant data samples planned for RHIC Runs 12 and 13.



\bibliographystyle{aipproc}   

\bibliography{sample}

\hyphenation{Post-Script Sprin-ger}
\begin{thebibliography}{8}
\expandafter\ifx\csname natexlab\endcsname\relax\def\natexlab#1{#1}\fi
\providecommand{\enquote}[1]{``#1''}
\expandafter\ifx\csname url\endcsname\relax
  \def\url#1{\texttt{#1}}\fi
\expandafter\ifx\csname urlprefix\endcsname\relax\def\urlprefix{URL }\fi
\providecommand{\eprint}[2][]{\url{#2}}

\bibitem{RHIC}
B.~I. Abelev et al. STAR Collaboration, {\it Phys. Rev. Lett.} {\bf 101}, 222001 (2008).

\bibitem{HERMES}
A.~Airapetian et al. HERMES Collaboration, {\it Phys. Rev. Lett.} {\bf 103}, 152002 (2009) [arxiv:0906.3918].

\bibitem{SignChange}
J.~C. Collins, {\it Phys. Lett. B} {\bf 536}, 43 (2002) [arXiv:hep-ph/0204004].

\bibitem{ANjet}
Z. Kang, J. Qiu, W. Vogelsang, F. Yuan, {\it Phys. Rev. D} {\bf 83}, 094001 (2011) [arxiv:1103.1591].

\bibitem{PAC2010}
http://www.bnl.gov/npp/docs/pac0610/Crawford\_LoI.100524.v1.pdf

\bibitem{PAC2011}
http://www.bnl.gov/npp/docs/pac0611/DY\_pro\_110516\_final.2.pdf


\end{thebibliography}

\IfFileExists{\jobname.bbl}{}
 {\typeout{}
  \typeout{******************************************}
  \typeout{** Please run "bibtex \jobname" to optain}
  \typeout{** the bibliography and then re-run LaTeX}
  \typeout{** twice to fix the references!}
  \typeout{******************************************}
  \typeout{}
 }

\end{document}